# *COBE*-DMR-normalisation for inflationary flat dark matter models

R. Stompor,[1] K. M. Górski[2,3] and A. J. Banday[4]
[1] *Copernicus Astronomical Center, Bartycka 18, 00-716 Warszawa, Poland, e-mail: radek@camk.edu.pl*
[2] *Universities Space Research Association, NASA/GSFC, Laboratory for Astronomy and Solar Physics Code 685, Greenbelt, MD 20771*
[3] *Warsaw University Observatory, Aleje Ujazdowskie 4, 00-478 Warszawa, Poland*
[4] *Hughes STX Corporation, 4400 Forbes Bldg, Lanham, MD 20706*

**ABSTRACT**

The two-year $COBE$-DMR 53 and 90 GHz sky maps, in *both* galactic and ecliptic coordinates, are used to determine the normalisation of inflationary universe models with a flat global geometry and adiabatic density perturbations. The appropriately normalised cold and mixed dark matter models and cosmological constant dominated, cold dark matter models, computed for a range of values of $\Omega_b$ and $h$, are then compared to various measures of structure in the universe. Critical density CDM models appear to be irreconcilable with observations on both large and small scales simultaneously, whereas MDM models provide a somewhat better fit to the data. Although the $COBE$-DMR data alone prefer a nearly critical value for the total density, low-density cosmological constant models with $\Omega_0 \geq 0.15$ can not be rejected at a confidence level exceeding 95%. Such models may also provide a significantly better fit to the matter distribution data than critical density CDM.

## 1 INTRODUCTION

The $COBE$-DMR discovery of cosmic microwave background (CMB) anisotropy (Smoot et al. 1992, Bennett et al. 1992, Wright et al. 1992) has affected cosmology in both epistemological and practical ways, but its predominant quantitative influence has been to provide the means for the accurate normalisation of theories of large scale structure formation.

The development of inflationary ideas during the 1980s induced over a decade-long adherence to the cosmological paradigm which posits that the universe is spatially flat. Such a picture requires that the present energy density of the universe is dominated by non-baryonic dark matter or alternatively by a non-zero vacuum energy contribution (a cosmological constant term, $\Omega_\Lambda$).

The minimal version of the model, which invokes cold dark matter (CDM) as the major constituent of the universe, is presently in direct confrontation with astronomical observations. Extensions of the model, which in addition to CDM postulate either an admixture of hot dark matter (HDM) or a non-zero cosmological constant (CDM-$\Lambda$), enjoy considerable popularity in contemporary cosmological research. Whilst vigorous discussion ensues in the literature as to the plausibility of the mixed dark matter (MDM) (see e.g. Schaefer, Schafi & Stecker 1989, Klypin et al. 1993, Pogosyan & Starobinsky 1993, Ma & Bertschinger 1994, Primack et al. 1995) and CDM-$\Lambda$ (e.g. Peebles 1984; Efstathiou, Maddox, & Sutherland 1990; Carrol, Press, & Turner 1992; Kofman, Gnedin, & Bahcall 1993; Krauss & Turner 1995; Ostriker & Steinhardt 1995) models as viable cosmologies, we consider both CDM and MDM versions of the inflationary scenario (with $\Omega_\Lambda = 0$) and CDM models with a non-zero cosmological constant.

In this paper, we use the linear angular power spectrum estimation technique (Górski 1994) to normalise dark matter models to the two year $COBE$-DMR anisotropy measurements (Bennett et al. 1994). Bunn, Scott & White (1995), Bunn & Sugiyama (1994) and Sugiyama (1994) have used different power spectrum normalisation schemes to obtain comparable results for those models in common with this paper. We then discuss some predictions for large-scale structure measures resulting from this normalisation.

## 2 CMB ANISOTROPY NORMALISATION PROCEDURE

### 2.1 Data Selection and Power Spectrum Inference Method

The $COBE$-DMR project generates sky maps from the time-ordered data in both galactic and ecliptic reference frames. Previously, published work by the DMR team has utilised only the galactic data sets, whilst the general community analysed the publically released sky maps in ecliptic co-ordinates. (Note that *both* sets of data are now available by anonymous ftp from the NSSDC). In this work, the two year $COBE$-DMR 53 and 90 GHz galactic sky maps are used identically as in Górski et al. (1994). In addition, we also utilise the ecliptic coordinate frame data sets as a check on the extent to which the coordinate dependent noise binning can affect the results of the analysis.

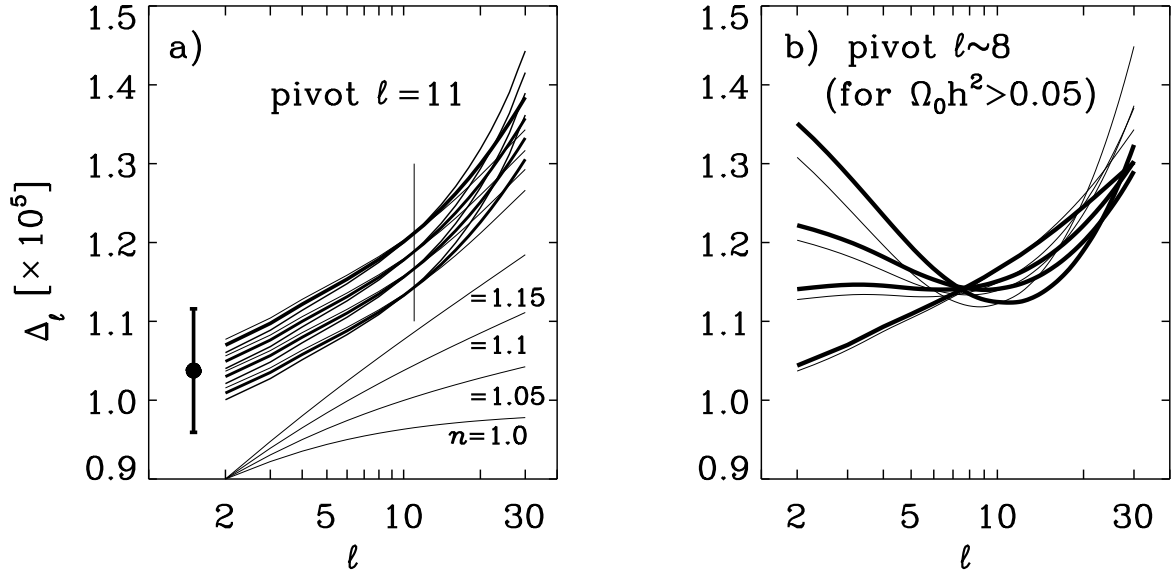

**Figure 1.** CMB anisotropy power spectra normalised to $COBE$-DMR, where $\Delta_\ell^2 = \ell(2\ell+1)a_\ell^2/4\pi$. *a)* CDM/MDM: the four sets of curves correspond to the power spectrum amplitude determination as follows (from top to bottom): 1) ecliptic coordinates, quadrupole excluded, 2) ecliptic coordinates, quadrupole included, 3) galactic coordinates, quadrupole excluded, and 4) galactic coordinates, quadrupole included. Within each set the three curves denote: $h = 0.5$ and $\Omega_b = 0.05$ - heavy type, $h = 0.3$ and $\Omega_b = 0.14$ - medium-heavy type, $h = 0.8$ and $\Omega_b = 0.02$ - light type, which are consistent with the constraints from BBN. The crossing point within each set is at $\ell \approx 11$. Several pure Sachs-Wolfe power law spectra are shown (shifted down to 0.9 at $\ell = 2$). Although a power law approximation, $P(k) \propto k^n$, used to generate the multipole coefficients $a_\ell^2$ solely through the Sachs & Wolfe (1967) effect will be poor for such spectra, a value of $n \simeq 1.1$ would be most appropriate over the range $\ell \lesssim 15$. This is a little steeper than the underlying, inflationary $n = 1$ spectrum (Bond 1993). The proposed overall two year $COBE$-DMR normalisation (translated from $Q_{rms-PS} \sim 20\,\mu K$) is represented by a filled circle; the error bar represents a typical statistical uncertainty of an individual likelihood fit. Clearly, this encompasses the fit uncertainties due to choice of coordinate system, quadrupole inclusion/exclusion, and/or cosmological parameter values.
*b)* CDM-$\Lambda$ : the thin lines correspond to the $h = 0.5, \Omega_b = 0.05$ models, thick lines to models with $h = 0.8$, $\Omega_b = 0.02$. The curves are drawn for four values of total density (in descending order at $\ell = 2$): $\Omega_0 = 0.1, 0.2, 0.3, 1$. The approximate pivot point (for $\Omega_0 h^2 \gtrsim 0.05$ models) is also shown.

We implement the power spectrum estimation method as described in considerable detail in Górski (1994). Co-ordinate system specific orthogonal basis functions for the Fourier decomposition of the sky maps are constructed so as to include exactly both pixelisation effects and a $|b| < 20°$ galactic plane excision (leaving 4016 and 4038 pixels in the galactic and ecliptic sky maps, respectively). A likelihood analysis is performed as in Górski et al. (1994).

### 2.2 Theoretical Spectra of Anisotropy

We consider spatially flat models with $\Omega_0 + \Omega_\Lambda = 1$ and primordial perturbations as prescribed (to within an arbitrary amplitude) by the inflationary scenario. Hence, random-phase, Gaussian, scalar curvature perturbations are assumed (no gravity waves), corresponding to adiabatic density perturbations with the Harrison-Zel'dovich power spectrum, $P(k) \propto k$.

For $\Omega_0 = 1$, the bulk mass density is provided by either cold or mixed dark matter; in the case of the MDM models, the hot dark matter is introduced in the form of either one or two (equal mass) families of massive neutrinos, with a contributed fraction of the critical density $\Omega_\nu = 0.15, 0.2,$ 0.25 and 0.3 for one massive flavour, and $\Omega_\nu = 0.2, 0.3,$ otherwise. For the CDM model, the values of the Hubble constant, $H_0 = 100\,h$ km s$^{-1}$ Mpc$^{-1}$, and baryon abundance are sampled at $h = 0.3, 0.4, 0.5, 0.6, 0.7,$ and 0.8, and $\Omega_b = 0.01, 0.03, 0.05, 0.07, 0.1$ (for all values of $h$), respectively; in order to trace the Big-Bang nucleosynthesis (BBN) relation, $\Omega_b = 0.013\,h^{-2}$ (Reeves 1994), we also use $\Omega_b = 0.02$ for $h = 0.8$, and $\Omega_b = 0.14$ for $h = 0.3$. For the MDM models, we choose $h = 0.5$ and $\Omega_b = 0.05$. The CDM-$\Lambda$ models considered are specified by either $h = 0.5$ or 0.8, and $\Omega_b$ is again selected to match BBN constraints, i.e. the baryon density is $\Omega_b = 0.05$ for $h=0.5$, and $\Omega_b = 0.02$ for $h = 0.8$. For comparison we also consider CDM-$\Lambda$ models with $\Omega_b = 0.03$ for both values of $h$.

The CMB anisotropy multipole coefficients and the matter perturbation transfer functions for both models were evaluated using the Boltzmann equation code of Stompor (1994) by solving the propagation equations up until the redshift $z = 0$. Thus all effects caused by the cosmological constant were taken into account exactly.

Fig. 1a shows a selection of the CDM radiation power spectra normalised to the $COBE$-DMR two year CMB anisotropy. Over the low-$\ell$ range of CMB multipoles probed by $COBE$-DMR the theoretical spectra are indistinguishable for $\Omega_0 = 1$ CDM and MDM models with equivalent $h$ and $\Omega_b$. In this case, the power spectrum amplitude derived from the data applies equally to both CDM and MDM mod-els. Fig. 1b corresponds to several CDM-$\Lambda$ radiation power spectra. At low-$\ell$ ($\lesssim 30$), the radiation power spectra are determined by the usual Sachs & Wolfe (1967, SW) effect and by two integrated SW effects (ISW), a low-redshift one – induced by the non-zero cosmological constant, (Kofman & Starobinsky 1985, Górski, Silk, & Vittorio 1992), and a high-redshift one – due to the non-negligible radiation energy density contribution to the total density after decoupling. For those $k$-modes which dominate the $a_\ell$ coefficients with $\ell \lesssim 10$, the high-redshift ISW effect simply cancels that part of the SW effect due to radiation density perturbations present on the last scattering surface. The sum of both effects therefore scales with cosmological parameters as would a matter perturbation generated potential. In particular, its dependence on $h$ is almost the same as that of the low-redshift ISW contributions, while the dependence on $\Omega_0$ differs because of the $\Omega_0$-dependence of the density perturbation growth rate at low redshifts. Hence, the shape of that part of the radiation transfer function depends very weakly on the Hubble constant. However, this is not the case over the $\ell$-range $10 \lesssim \ell \lesssim 30$, where the canceling of the high-$z$ ISW and "radiation" part of the SW effect is less complete and the dependence of the power spectrum on $h$ is clearly noticeable. The strength of both ISW effects falls with increasing $\Omega_0$. Consequently, with respect to the fiducial pure Harrison-Zel'dovich form $\sim 1/l(l+1)$, the power spectra flatten as $\Omega_0$ increases, and for models with near critical matter density the spectra monotonically grow with $\ell$. This steepening of the low-$\ell$ end of the power spectrum over the pure SW case is caused by the high-redshift ISW effect, which, though small, does not vanish even in critical density cases (Bond 1993; Bunn et al. 1995), and a small contribution from the first "Doppler" peak. The steepest spectra arise in models near critical matter density ($\Omega_0 \simeq 0.8$).

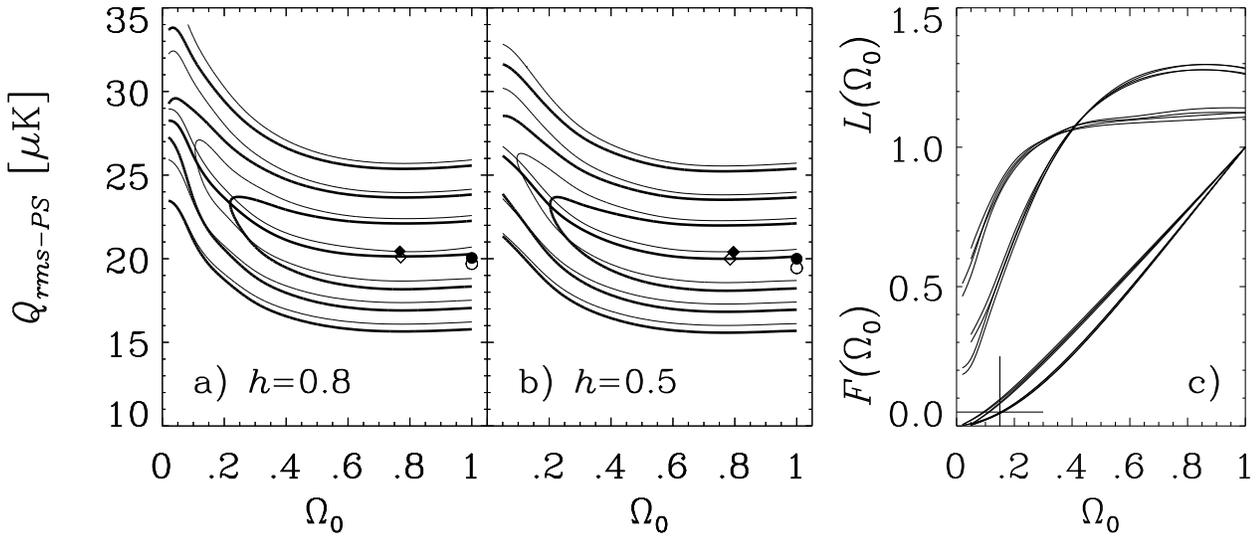

**Figure 2.** *a)* Contour plots of the likelihood density in the $Q_{rms-PS} - \Omega_0$ plane for the $h = 0.8$ CDM-$\Lambda$ models determined from analysis of the two year *COBE*-DMR data in ecliptic coordinates. 68%, 95%, and 99.8% integrated likelihood levels, derived under the assumption of a uniform $(Q_{rms-PS}, \Omega_0)$–prior, and the $\Omega_0$ =const–ridge lines are shown. Thick lines correspond to the results of the analysis with the quadrupole included, thin lines excluding the quadrupole. Diamonds show the loci of the likelihood maxima: filled symbols show the situation when the quadrupole is included, open symbols when it is excluded. The filled and open circles show the likelihood maxima loci derived from analysis of the galactic data to illustrate the small shift in the inferred model amplitude. *b)* As in *a* for $h = 0.5$.
*c)* Marginal likelihood densities for $\Omega_0$, derived by integration of the two dimensional density distributions (from panels *a* and *b*) over $Q_{rms-PS}$, are shown by thin lines. The functions corresponding to the analysis with the quadrupole included fall more steeply with decreasing $\Omega_0$ than when the quadrupole is excluded. The lines for the $h = 0.8$ model extend down to $\Omega_0 = 0.02$. Results derived with both the ecliptic and galactic frame data are shown — the ecliptic frame results fall below the galactic ones at small $\Omega_0$. The heavy lines show the cumulative likelihood functions derived from the above likelihood densities. The quadrupole included case renders more steeply rising curves with $h = 0.5$, 0.8 models, and galactic/ecliptic results all merge together. The less steeply rising curves, split at very small values of $\Omega_0$, correspond to the no quadrupole case. The 5% cumulative likelihood level is shown, as is the overall most restrictive, $h$- and coordinate frame-independent, $\Omega_0 \geq 0.15$ limit on the CDM-$\Lambda$ models implied by the two year *COBE*-DMR data alone.

We parameterise all such spectra by the exact value of the quadrupole, $Q_{rms-PS}$ (the value of the exactly computed quadrupole, $a_2$, multiplied by $(5/4\pi)^{1/2}$, which depends non-trivially on cosmological parameters), in a straightforward generalization of the $Q_{rms-PS}$ introduced in Smoot et al. (1992) for pure power law model spectra. When $\Omega_0 \lesssim 0.5$, the ISW contributions boost the quadrupole over its pure Sachs & Wolfe counterpart. For $\Omega_0 \gtrsim 0.5$ — models with a negligible cosmological constant induced ISW effect — the resulting exact quadrupole is lower than its pure SW value (see Table 2). In the case of critical density models, these differences, mainly due to the high-redshift, integrated SW effect, are rather small, being equal to $\sim 10\%$, 5%, or 2% for $h \lesssim 0.4$, $h \sim 0.5$, and $h \sim 0.8$, respectively, of the pure SW value.

### 2.3 Results of $Q_{rms-PS}$ Fitting

A typical likelihood fit of flat dark matter models to the two year $COBE$-DMR data yields a $\sim 13\sigma$ significant determination of $Q_{rms-PS}$, with a value of $\sim 20\,\mu$K for the critical matter density case, and in the range $\sim 20 - 26\,\mu$K for $\Omega_\Lambda > 0$. Systematic shifts in the central value of the fit are observed due to: 1) differences in the noise pixelisation in the galactic and ecliptic coordinate frames, which result in a $\sim 0.8\,\mu$K difference between the inferred normalisation amplitudes, with higher values obtained from the ecliptic maps (an effect alluded to, but not quantified, by Bunn et al. 1995); 2) exclusion of the quadrupole, which produces a $\sim +0.4\,\mu$K variation in the fitted amplitude; 3) the uncertainty in the values of $h$ and $\Omega_b$ (reflected by differences in spectral shape over the $\ell$-range accessible to DMR) causes an additional small spread, $\pm 0.25\,\mu$K, of the fitted amplitudes.

Likelihood contours in the $Q_{rms-PS} - \Omega_0$ plane (under the assumption of a uniform prior) are shown in Figs. 2a, b. The preferred value of the matter density is clearly close to critical, but even low-density models with $\Omega_0 \geq 0.15$ lie well within the 95% confidence limits and appear acceptable. Due to the weak dependence of the $COBE$-DMR normalised $Q_{rms-PS}$ on $h$, this 95% C.L. lower limit is essentially Hubble constant independent.

A convenient summation of the proposed overall normalisations for the flat, dark matter models is then:

$$Q_{rms-PS}(\mu K) \simeq [20 + g(\Omega_0, h)] \times (1 \pm 0.077)$$
$$\pm 0.4 \quad \pm 0.2 \quad \pm 0.25$$

where $g(\Omega_0, h) = 9.1 h^{0.6} \exp(-\Omega_0^2 h^{0.3}/0.057)$ for $\Omega_0 \leq 1$, and $g(\Omega_0, h) = 0$ for $\Omega_0 = 1$. The error ranges represent the statistical error and uncertainties associated with effects 1 through 3 above, respectively. It will be noted that the statistical error on the inferred normalisation is considerably larger than the other uncertainties. Exact values for a representative set of CDM-$\Lambda$ models can be found in Table 2, which can be easily interpolated to other models with high accuracy. However, our fit is good to within 5%.

Górski et al. (1994) showed that for the power law models specified by $Q_{rms-PS}$ and the spectral index $n$ a convenient, $n$-independent normalisation was in terms of the multipole amplitude $a_9 \simeq 8\,\mu$K. This $\ell$-order is related to the point at which the theoretical signal to noise ratio is of order unity. For the $\Omega_0 = 1$ dark matter models, we find that the appropriate pivot point is at $a_{11} \simeq 7.15 \pm 0.55\mu$K. However, those models with a non-zero cosmological constant show no pivot point at a common multipole order. Nevertheless, for those models with $\Omega_0 h^2 \gtrsim 0.05$, a good approximate pivot can be found at $l \simeq 8$ (Fig. 1b), yielding the model-independent normalisation $a_8 \simeq 9.5 \pm 0.73\mu$K.

### 3  MEASURES OF LARGE-SCALE STRUCTURE

Having determined the normalisations for our grid of models, we now proceed to discuss the values of several large-scale structure statistics. These have been computed from the matter perturbation spectra according to the usual prescriptions given as footnotes to Table 1 (which contains representative values for the $\Omega_0 = 1$, $h = 0.5$, $\Omega_b = 0.05$ models). Table 2 summarises the pertinent values for a set of cosmological constant models.

### 3.1  Mass Fluctuations: $\sigma_8$ and $J_3$

$COBE$-DMR normalised values of $(\sigma_8)_{mass}$ are shown in Fig. 3a. A related, observable quantity is the fluctuation in the number of galaxies within a sphere of fixed radius. Recent estimates are close to the standard value from Davis & Peebles (1983), $(\sigma_8)_{gal} \simeq 1$ — a representative selection of the galaxy surveys is: $(\sigma_8)_{gal} \simeq 0.8$ (Fisher et al. 1993), $0.87 \pm 0.07$ (Feldman, Kaiser & Peacock, 1994), and $0.83^{+0.05}_{-0.07}$ (Baugh & Efstathiou, 1993).

In order to compare the rms mass fluctuations with the rms galaxy fluctuations, we adopt the simple picture of biased galaxy formation, wherein $(\sigma_8)_{gal} = b\,(\sigma_8)_{mass}$, and $b$ is the linear bias factor. Some constraints on the value of $b$ can be imposed by recent galaxy surveys, but are more uncertain than the $(\sigma_8)_{gal}$ determinations and usually inseparable from the total density parameter.

In the $\Omega_0 = 1$ case, the estimated values of $b$ range from 0.9 to 2, with strong bimodal behaviour about $b \sim 1$ and 2 (see Table 1 in Dekel 1994, and Table 1 of Cole, Fisher & Weinberg 1994, with $\Omega_0 = 1$). Estimates of $(\sigma_8)_{mass}$ inferred from galaxy cluster catalogues favour a higher value for $b$. Henry & Arnaud (1991) used the abundance of clusters as a function of X-ray temperature to derive $(\sigma_8)_{mass} = 0.59 \pm 0.02$ for a scale-free, flat universe. White, Efstathiou & Frenk (1993) have used the masses and abundances of rich clusters of galaxies to determine $(\sigma_8)_{mass} \simeq 0.52 - 0.62$ for a critical density universe (relatively independent of $h$). A low value of $(\sigma_8)_{mass} \lesssim 0.5$ is also required in critical density models by the observed low pair-wise velocities (Davis et al. 1985, but see Zurek et al. 1994). On the other hand, Seljak & Bertschinger (1994) determine a 95% confidence range for $(\sigma_8)_{mass}$ of 0.7 - 2.3 from the POTENT density field reconstruction.

**Table 1.** Inferred cosmological statistics for models with $\Omega_0 = \Omega_{CDM} + \Omega_\nu + \Omega_b = 1$, $\Omega_b = 0.05$, $h = 0.5$, and a $COBE$-DMR normalisation of $Q_{rms-PS} = 20.04\,\mu$K. $N_\nu$ is the number of massive neutrino species, and $m_\nu$ the neutrino mass in eV. The errors, including both statistical (1$\sigma$) and systematic deviations, are of the order of 11%.

| $\Omega_\nu$ | $N_\nu$ | $m_\nu$ | $(\sigma_8)_{mass}^{(a)}$ | $J_3(20)^{(b)}$ | $v_{40}^{(c)}$ | $v_{60}^{(c)}$ | $v_{100}^{(c)}$ |
|---|---|---|---|---|---|---|---|
| 0.00 | – | 0.0 | 1.36 | 968 | 444 | 355 | 248 |
| 0.15 | 1 | 3.7 | 0.97 | 706 | 441 | 357 | 251 |
| 0.20 | 1 | 4.9 | 0.92 | 694 | 442 | 358 | 252 |
| 0.25 | 1 | 6.1 | 0.88 | 691 | 444 | 359 | 252 |
| 0.30 | 1 | 7.3 | 0.85 | 695 | 445 | 360 | 252 |
| 0.20 | 2 | 2.4 | 0.82 | 567 | 435 | 356 | 252 |
| 0.30 | 2 | 3.7 | 0.71 | 516 | 439 | 359 | 254 |

(a) $(\sigma_{hR})^2_{mass} = \frac{1}{2\pi^2} \int_0^\infty w_{TH}^2(kR)\,P(k)k^2\,dk$;

(b) $J_3(hR) = \frac{R^3}{2\pi^2} \int_0^\infty w_{TH}(kR)\,P(k)\,k^2\,dk$, in $(h^{-1}\mathrm{Mpc})^3$, and for $R = 20 h^{-1}\mathrm{Mpc}$;

(c) $v_{hR}^2 = \frac{1}{2\pi^2} \int_0^\infty w_{TH}^2(kR)\,e^{-k^2 r_s^2}\,P_v(k)k^2\,dk$, $hr_s = 12\mathrm{Mpc}$, in km s$^{-1}$;

where $w_{TH}(x) = 3 j_1(x)/x$,

$P(k)$ and $P_v(k)$ denote the present-day density and velocity perturbation power spectra, respectively.

**Table 2.** Inferred cosmological statistics for chosen spatially flat CDM-$\Lambda$ models with $\Omega_b = 0.013h^{-2}$, and a *COBE*-DMR normalisation, expressed in terms $Q_{rms-PS}$. The errors, including both statistical ($1\sigma$) and systematic deviations, are of the order of 11%.

| $\Omega_0$ | $h$ | $Q_{rms-PS}^{(a)}$ | $(\sigma_8)_{mass}$ | $J_3^{(b)}$ | $v_{40}^{(c)}$ | $v_{60}^{(c)}$ | $v_{100}^{(c)}$ | $a_2^2/a_{2(sw)}^{2\ (d)}$ |
|---|---|---|---|---|---|---|---|---|
| 0.02 | 0.8 | 28.16 | 0.04 | 4 | 52 | 51 | 50 | 2.64 |
| 0.05 | 0.8 | 27.88 | 0.21 | 66 | 117 | 113 | 105 | 2.29 |
|  | 0.5 | 26.09 | 0.05 | 7 | 78 | 76 | 73 | 1.98 |
| 0.10 | 0.8 | 26.11 | 0.53 | 335 | 208 | 192 | 167 | 1.90 |
|  | 0.5 | 25.27 | 0.17 | 41 | 139 | 134 | 123 | 1.64 |
| 0.20 | 0.8 | 23.61 | 1.03 | 985 | 330 | 290 | 230 | 1.46 |
|  | 0.5 | 23.24 | 0.44 | 222 | 235 | 216 | 183 | 1.28 |
| 0.30 | 0.8 | 22.04 | 1.39 | 1496 | 403 | 342 | 258 | 1.22 |
|  | 0.5 | 21.79 | 0.66 | 428 | 301 | 267 | 214 | 1.10 |
| 0.40 | 0.8 | 21.10 | 1.66 | 1825 | 449 | 372 | 272 | 1.09 |
|  | 0.5 | 20.93 | 0.84 | 607 | 349 | 301 | 233 | 0.99 |
| 0.60 | 0.8 | 20.20 | 1.99 | 2081 | 497 | 398 | 279 | 0.97 |
|  | 0.5 | 20.07 | 1.09 | 842 | 405 | 338 | 249 | 0.90 |
| 0.80 | 0.8 | 20.02 | 2.17 | 2033 | 513 | 403 | 276 | 0.94 |
|  | 0.5 | 19.91 | 1.26 | 950 | 433 | 353 | 252 | 0.89 |
| 1.00 | 0.8 | 20.17 | 2.25 | 1859 | 515 | 399 | 269 | 0.96 |
|  | 0.5 | 20.04 | 1.36 | 968 | 444 | 355 | 248 | 0.91 |

[a] in $\mu$K;

[b] in $(h^{-1}\mathrm{Mpc})^3$, and for $R = 20h^{-1}\mathrm{Mpc}$;

[c] in km s$^{-1}$.

[d] $a_{2(sw)}$ is the pure Sachs-Wolfe contribution to the quadrupole.

Although the value of the rms mass fluctuations is not yet well known, it seems unlikely that the biasing parameter is less than unity. The generally high values of $(\sigma_8)_{mass} > 1$ predicted from the *COBE*-DMR normalised $\Omega_0 = 1$, CDM models are difficult to reconcile with observations. Physical motivations for requiring a bias parameter somewhat larger than unity were pointed out early on (e.g. Silk 1977), and are supported by recent hydrodynamical simulations (Katz, Hernquist & Weinberg 1992, Cen & Ostriker 1992) which suggest a value $b \sim 1.5$. The values of $b \geq 1$ would require either pure CDM in an $\Omega_0 = 1$ universe with a low-$h$ or high-$\Omega_b$, in disagreement with observations ($0.4 \lesssim h \lesssim 0.8$) or the BBN constraints, respectively, or alternatively an appreciable admixture of a hot dark component.

A non-vanishing cosmological constant can also provide a means to circumvent these problems. The predicted $(\sigma_8)_{mass}$ as a function of $\Omega_0 h^2$ is shown in Fig. 3b. A low value of $\Omega_0$ suppresses peculiar motions so high bias is not required to fit the small-scale galaxy pair-wise velocities. Thus unbiased models with $(\sigma_8)_{mass} \simeq (\sigma_8)_{gal}$ are likely to be of interest.

Note that a high biasing parameter ($b \gtrsim 1.5$) together with the $(\sigma_8)_{mass} \simeq 0.57\Omega_0^{-0.56}$ (White et al. 1993, Strauss & Willick 1995), estimation based on galaxy cluster catologs would require an almost critical density model with a low ($h \lesssim 0.3$) Hubble constant in order to be compatible with the *COBE*-DMR measurement. The situation is even more dramatic for such models if the Seljak & Bertschinger (1994) estimation of rms mass fluctuations, $(\sigma_8)_{mass} \simeq 1.3^{+1.0}_{-0.6}\Omega_0^{-0.6}$, is considered. In this case, none of the highly biased ($b \gtrsim 1.5$) subcritical total density models can fulfill this constraint, and even the relaxed requirement that $b \geq 1$ imposes a 95% C.L. bottom limit on the total density $\Omega_0 \gtrsim 0.55$ which, in conjunction with the *COBE*-DMR data, again imposes an upper limit on $h$ ($\lesssim 0.5$). Until the apparent discrepancy between the two $(\sigma_8)_{mass}$ estimates is resolved, a conservative approach is adopted with $b \geq 1$ as suggested above, so that models with $\Omega_0 h^2 \lesssim 0.13$ are also relevant.

Comparison of the $J_3$-integral, computed on the larger scale of $R = 20h^{-1}$Mpc (which, in principle, is less contaminated by non-linear galaxy power spectrum evolution than $\sigma_8$), to the observational value from Davis & Peebles (1983), $J_3(20h^{-1}\mathrm{Mpc}) \simeq 700\, h^{-3}\, \mathrm{Mpc}^3$ (probably accurate to within $\sim 30\%$), also seems to favour the low-$\Omega_0$ CDM-$\Lambda$-models (Fig. 4) in agreement with the above constraint.

### 3.2 Large-Scale Flows

The local streaming motions of galaxies provide an interesting constraint for cosmological models. In particular, galaxy peculiar velocities directly measure mass fluctuations, independently of a linear bias parameter. Dekel (1994) gives estimates of the average peculiar velocities within spheres of radius 1000 to 6000 km s$^{-1}$, which seem to be in agreement with the HI data from Giovanelli & Haynes (Dekel, private communication). The *COBE*-DMR normalised rms bulk flows are shown in Fig. 5. The critical density model predictions are in good agreement with the POTENT data, and it is apparent that all models with $\Omega_0 h \gtrsim 0.06$ can account for the POTENT velocities. However, an important recent development in the field was the determination by Lauer & Postman (1994) of significant bulk flow in a deep volume limited sample of $\sim 100$ galaxy clusters. If confirmed, even the critical density models will be in serious trouble (since all of the presently considered models of structure formation predict too rapid a decrease in the bulk flow amplitude with scale. See also Feldman & Watkins 1994,

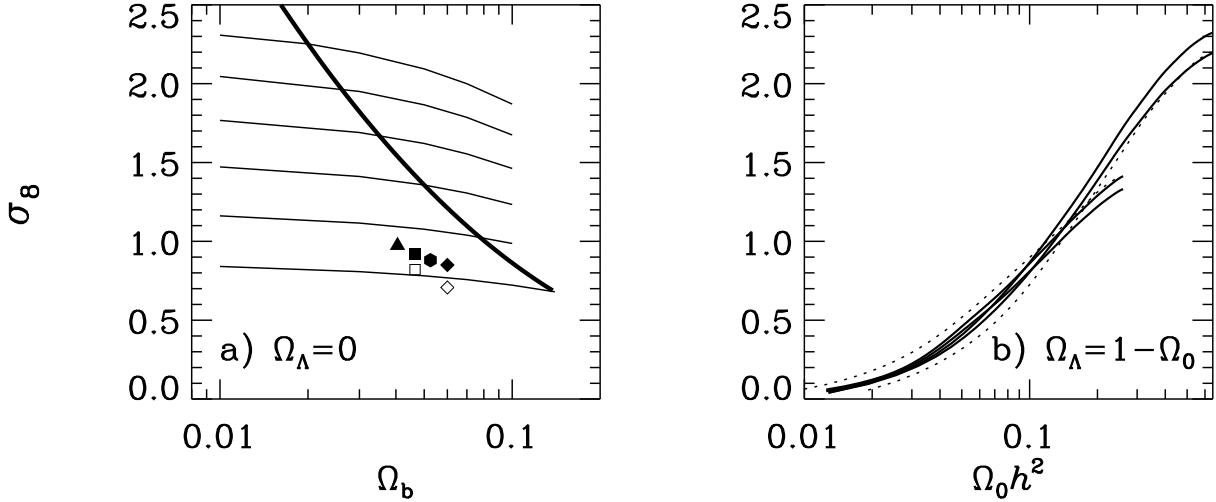

**Figure 3.** *a)* $(\sigma_8)_{mass}$ values predicted from the *COBE*-DMR normalised, flat, dark matter dominated models. The thin solid lines correspond to the CDM model with $h$ decreasing from 0.8 to 0.3 in steps of 0.1 from top to bottom. The thick solid line shows the models which obey the Big Bang nucleosynthesis constraint, $\Omega_b = 0.013\,h^{-2}$. The individual points represent several mixed dark matter models: triangle - $\Omega_\nu = 0.15$, square - $\Omega_\nu = 0.20$, hexagon - $\Omega_\nu = 0.25$, diamond - $\Omega_\nu = 0.30$, where the filled symbols have $N_\nu = 1$, open symbols $N_\nu = 2$. All of the points correspond to $h = 0.5$ and $\Omega_b = 0.05$ but are spread out on the plot for clarity.
*b)* $(\sigma_8)_{mass}$ values predicted from the *COBE*-DMR normalised, CDM-$\Lambda$ models, plotted as a function of $\Omega_0 h^2$. The $h = 0.5$, $\Omega_b = 0.05$ models are shown by solid lines terminating at $\Omega_0 h^2 = 0.25$, while the $h = 0.8$, $\Omega_b = 0.02$ lines extend to the edge of the plot. For each model, the upper of the pair of lines corresponds to the ecliptic frame/no quadrupole analysis, whilst the lower represents the galactic frame/quadrupole included case (the remaining two combinations would lie in between). Dotted lines show the $\Omega_b = 0.03$ results averaged over the four possible combinations.

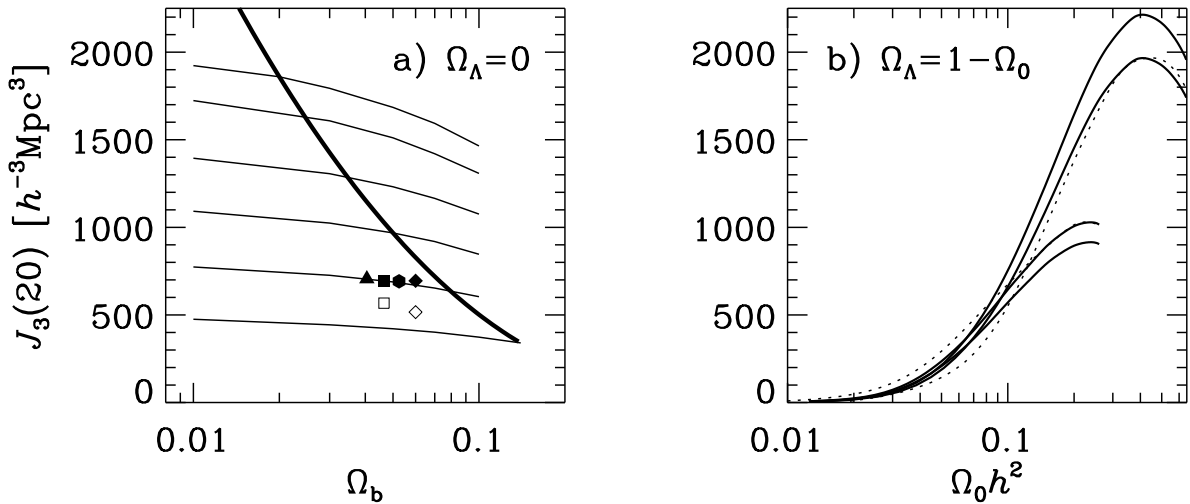

**Figure 4.** *a)* The values for the $J_3$ integral over the density perturbation correlation function within 20 $h^{-1}$Mpc (units $[h^{-1}\mathrm{Mpc}]^3$). Same coding as 3 *a*.
*b)* The values for the $J_3$ integral over the density perturbation correlation function within 20 $h^{-1}$Mpc (units $[h^{-1}\mathrm{Mpc}]^3$). Same coding as 3 *b*.

Strauss et al. 1995.).

### 3.3 Galaxy and Matter Distribution Power Spectra

In principle, the most direct and informative comparison of theory and observations can be conducted using the predicted and empirical power spectra. Several recently estimated galaxy power spectra are shown in Fig. 6, together with various theoretical linear power spectra (assuming $b$ =1). However, such a comparison is still subject to difficulties, including the linear nature of our theoretical computa-tions (since on small scales, the effects of non-linear evolution should be apparent), possible distortions introduced by redshift-real space mapping, and the lack of agreement between different observational estimates of the galaxy power spectrum (which may be explicable in terms of intrinsic variations in the survey samples and/or by differences in the estimation technique utilised). In spite of such difficulties, we still consider such a comparison to be informative in a qualitative way.

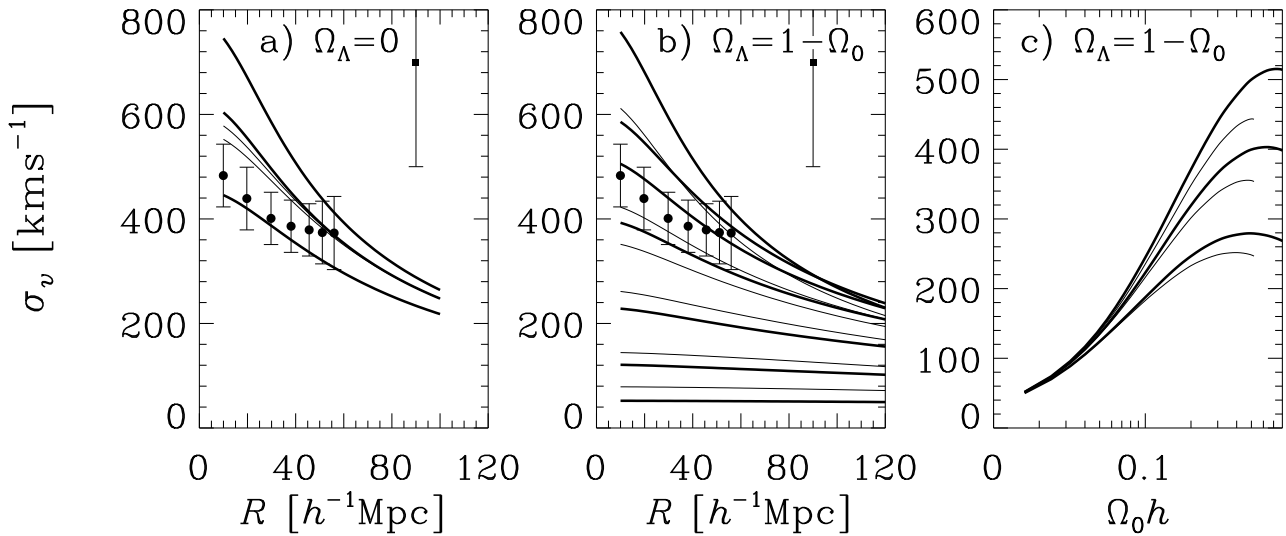

**Figure 5.** *a)* rms amplitudes of the large scale flows ([km s$^{-1}$]). The heavy lines correspond to the CDM models with $h = 0.8$, $\Omega_b = 0.02$ — top, $h = 0.5$, $\Omega_b = 0.05$ — middle, $h = 0.3$, $\Omega_b = 0.14$ — bottom. Thin lines separating from the $h = 0.5$ curve correspond to the MDM models with $N_\nu = 1$ (higher line), and $N_\nu = 2$, practically independent of $\Omega_\nu$. POTENT data (courtesy A. Dekel) are shown by circles, and the square shows the Lauer & Postman datum.
*b)* rms amplitudes of the large scale flows ([km s$^{-1}$]) as a function of the top-hat sphere radius encompassing the sample (an additional $r_s = 1200$ km s$^{-1}$ smoothing is included). Heavy lines — $h = 0.8$, $\Omega_b = 0.02$, and $\Omega_0 = 1$, 0.4, 0.3, 0.2, 0.1, 0.05, and 0.02 in descending order. Thin lines — $h = 0.5$, $\Omega_b = 0.05$, and $\Omega_0 = 1$, 0.4, 0.3, 0.2, 0.1, and 0.05 in descending order.
*c)* rms amplitudes of the bulk flow ([km s$^{-1}$]) within spheres of 40, 60, and 100 $h^{-1}$Mpc plotted in descending order, respectively, as a function of $\Omega_0 h$. Heavy lines — $h = 0.8$, $\Omega_0 = 0.02$, thin lines — $h = 0.5$, $\Omega_0 = 0.05$.

It is apparent from Fig. 6a that none of the empirical real space power spectra (Baugh & Efstathiou 1993, Peacock & Dodds 1994) exhibit as much small-scale power as the linear *COBE*-DMR normalised $\Omega_0 = 1$ CDM spectrum. (Note, however, that the redshift space spectra from da Costa et al. (1994) indicate more power on small scales than the real space APM spectrum, a discrepancy which can only become more pronounced when one recalls that the real-to-redshift space mapping suppresses power on scales $k \gtrsim 0.1\,h$Mpc$^{-1}$ – Gramann, Cen & Bahcall 1993 – and would require high, scale dependent bias.) This could be considered as supportive of the MDM models. However, the high amplitude of the matter power spectra implied by the *COBE*-DMR measurements leaves very little room for the biasing parameter to exceed unity, in conflict with the $b$-estimates from cluster properties. Baugh & Efstathiou (1993) have suggested that, with $b \sim 2$, the matter power spectrum required by the observed galaxy distribution should look rather like that for the $\Omega_0 = 1$, $\Gamma = 0.2$ model (here, we follow the parametrization of the matter power spectra introduced by Efstathiou, Bond, & White 1992) depicted in Fig. 6a as a thin solid line. Alas, *none* of the *COBE*-DMR-normalised CDM or MDM model power spectra resembles such an *ad hoc* spectrum.

Considering the non-zero Λ models shown in Fig. 6b, we see that on small scales ($k > 0.1\,h$Mpc$^{-1}$), the estimated power spectra from Baugh & Efstathiou (1994), Peacock & Dodds (1994) and the IRAS 1.2 Jy redshift survey are in reasonable agreement for those models with $\Omega_0 h^2 \lesssim 0.08$. This limit is somewhat more restrictive than that obtained previously, but, combined with the *COBE*-DMR plus POTENT inferred lower limits on $\Omega_0$, still leaves room for acceptable models. (If the more stringent gravitational lensing constraints on $\Omega_\Lambda < 0.7$ – Maoz & Rix 1993 – are taken into account, this range shrinks somewhat). However, since apparent discrepancies exist between different empirical power spectra on these scales which can not be solely ascribed to redshift space induced distortions, relaxation of the above constraint would seem to be in order until the observational situation is clarified. Conservatively, we retain the previous limit.

However, it then appears that the preferred CDM-Λ models can not simultaneously fit the large-scale behaviour indicated by the APM data. This is more properly described by an unbiased CDM $h = 0.5$ model, which, unfortunately, fails noticeably on smaller scales. One should note, however, that on these larger scales there is again pronounced disagreement between different data sets (with the SSRS2+CfA2 data of da Costa et al., 1994, clearly overshooting all other estimates by a factor $\sim 4$, which cannot be completely explained by the effect of projecting the power spectrum to redshift space. Hence, it would require $b \lesssim 2\Omega_0^{0.56}/5 < 1$ – Kaiser 1987. For values of $b$ in the range $\simeq 1 - 2$, the linear correction appropriate for scales $k \lesssim 0.03\,h$Mpc$^{-1}$ – Gramann et al. 1993 – is only 10%–30% for $\Omega_0 = 0.1$ and 20%-50% for $\Omega_0 = 0.4$ so the pronounced power enhancement survives.). It undoubtedly remains too early for any definitive conclusions to be reached. If the SSRS2+CfA2 data are confirmed, the power enhancement visible on scales $k \sim 0.01\,\mathrm{Mpc}^{-1}$ may favour the low-$\Omega_0$ CDM-$\Lambda$ models. Improved, very deep galaxy catalogs may provide us with a definitive test of CDM-$\Lambda$ models.

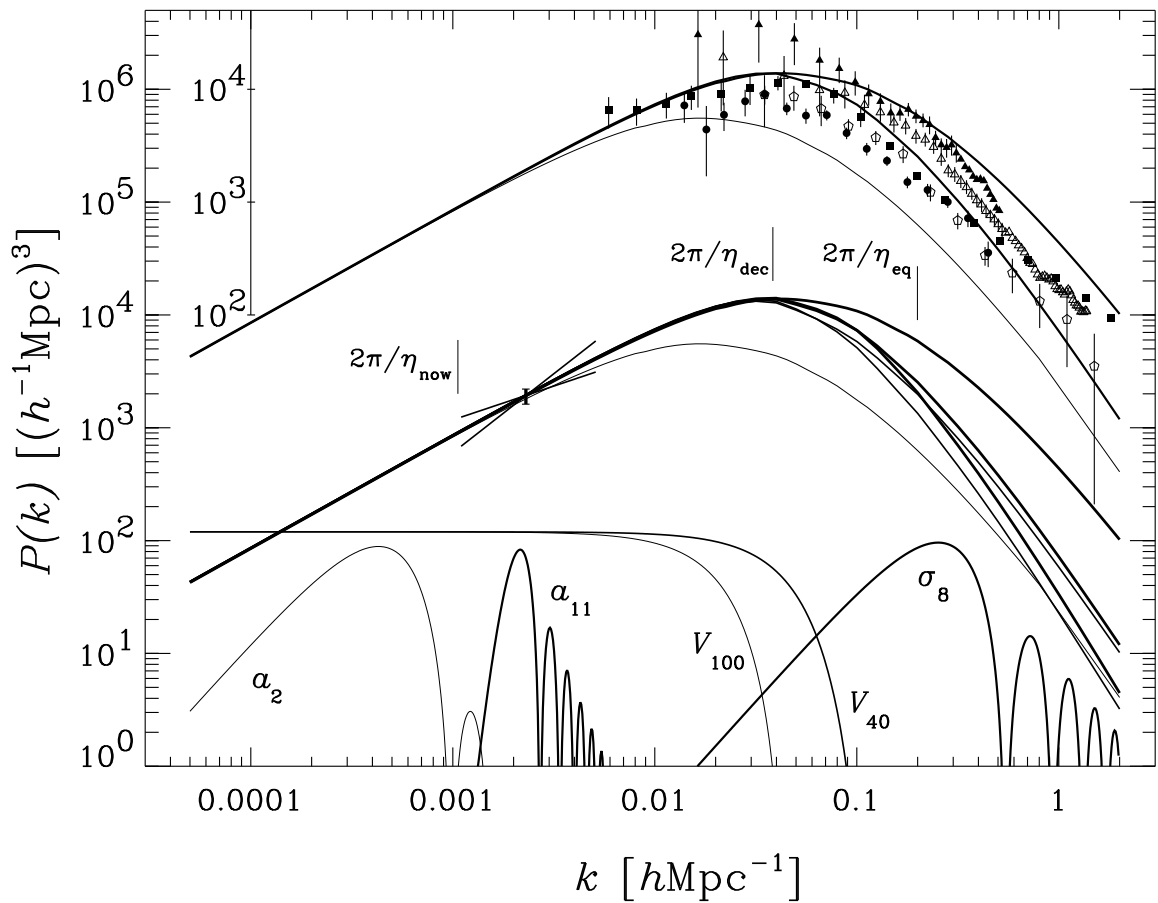

**Figure 6a.** *COBE*-DMR normalised inhomogeneity power spectra and, shown as the lowermost curves, miscellaneous spectral windows required for the computation of the statistics considered in this paper (as annotated). Overplotted are several observational estimates for the galaxy distribution power spectrum: squares - Baugh & Efstathiou (1993), filled circles - Peacock & Dodds (1994), pentagons - Fisher et al. (1993), open and filled triangles - two data subsets from da Costa et al. (1994). The theoretical mass distribution power spectra for the $\Omega_0 = 1$, $h = 0.5$, $\Omega_b = 0.05$ models are shown in the middle section of the plot: the top heavy line shows the CDM power spectrum; the lower heavy lines correspond to two $N_\nu = 1$ MDM models — $\Omega_\nu = 0.2$ (upper) and $\Omega_\nu = 0.3$ (lower); the medium heavy lines correspond to the equivalent $N_\nu = 2$ MDM models. Note that the MDM transfer function for $N_\nu = 2$ drops faster near $k \sim 0.04\,h\mathrm{Mpc}^{-1}$ $N_\nu = 1$. The thin line shows the $\Gamma = 0.2$ transfer function for CDM. The vertical error bar on the $P(k) \sim k$ part of the spectrum above the $a_{11}$ window illustrates the $\sim 1\,\sigma$ allowed variation in amplitude for this fixed slope. Conversely, the superposed 'bow' attempts to indicate the allowed $1\,\sigma$-variation ($\sim 0.3$) in tilt for spectra with the fixed $a_{11} = 7.15\,\mu$K amplitude. Unfortunately, there is no obvious way to represent the two dimensional uncertainty in the spectrum determination from the two year *COBE*-DMR data for flat dark matter models. One is *not* at liberty to arbitrarily vary both amplitude and spectral slope simultaneously. This is equivalent to the observed degeneracy seen in the two-dimensional ($Q_{rms-PS}$, $n$) fits to pure power law models (Górski et al. 1994 and references therein). The upper section of the plot reproduces the theoretical spectra for CDM, MDM ($N_\nu = 1$, $\Omega_\nu = 0.2$), and $\Gamma = 0.2$-CDM – all shifted upward for clarity.

## 4 DISCUSSION

The improved quality of the two year *COBE*-DMR data combined with reliable power spectrum estimation techniques allows the accurate normalisation of cosmological theories. Previously, the variance of *COBE*-DMR temperature fluctuations on a 10° angular scale was utilised for the normalisation of the power spectrum. Subsequent work (Wright et al. 1994, Banday et al. 1994) has demonstrated that this technique can be unreliable without considerable attention. More appropriate methods take full advantage of the measured CMB anisotropy power distribution on all angular scales accessible to the *COBE*-DMR instrument, as implemented in this paper.

The cold dark matter theory with a standard choice of cosmological parameters requires a very high normalisation in order to fit the CMB anisotropy distribution. Analysis of the first year of *COBE*-DMR data had already suggested that $(\sigma_8)_{mass} \sim 1$ (Wright et al. 1992, Efstathiou et al. 1992), and this value increases to $\sim 1.4$ with two years of data and an improved analysis technique. Although this normalisation allows the theory to predict large-scale velocities of comfortably high amplitude, it also results in a significant overproduction of density perturbations on scales $\lesssim 20\ h^{-1}$Mpc. CDM has often been criticised for its poor match to both galaxy and cluster distributions. Mixed dark matter models manage to circumvent, to a certain degree, these same problems by construction — massive neutrinos partially damp the perturbations at those length-scales where CDM looks problematic. Among the MDM models those with two species of massive neutrino seem to meet the observational constraints more comfortably (see also Primack et al. 1995). The larger free-streaming radius allows for the suppression of the perturbation amplitude on larger scales than in models with one massive flavour. This is reflected in the decrease of the predicted $(\sigma_8)_{mass}$ values. Nevertheless, the proponents of MDM will have to address the viability of the model viz. the simultaneous requirements that there should be no bias between the galaxy and mass distribution (as suggested by this analysis) and the galaxy pair-wise velocities should be small.

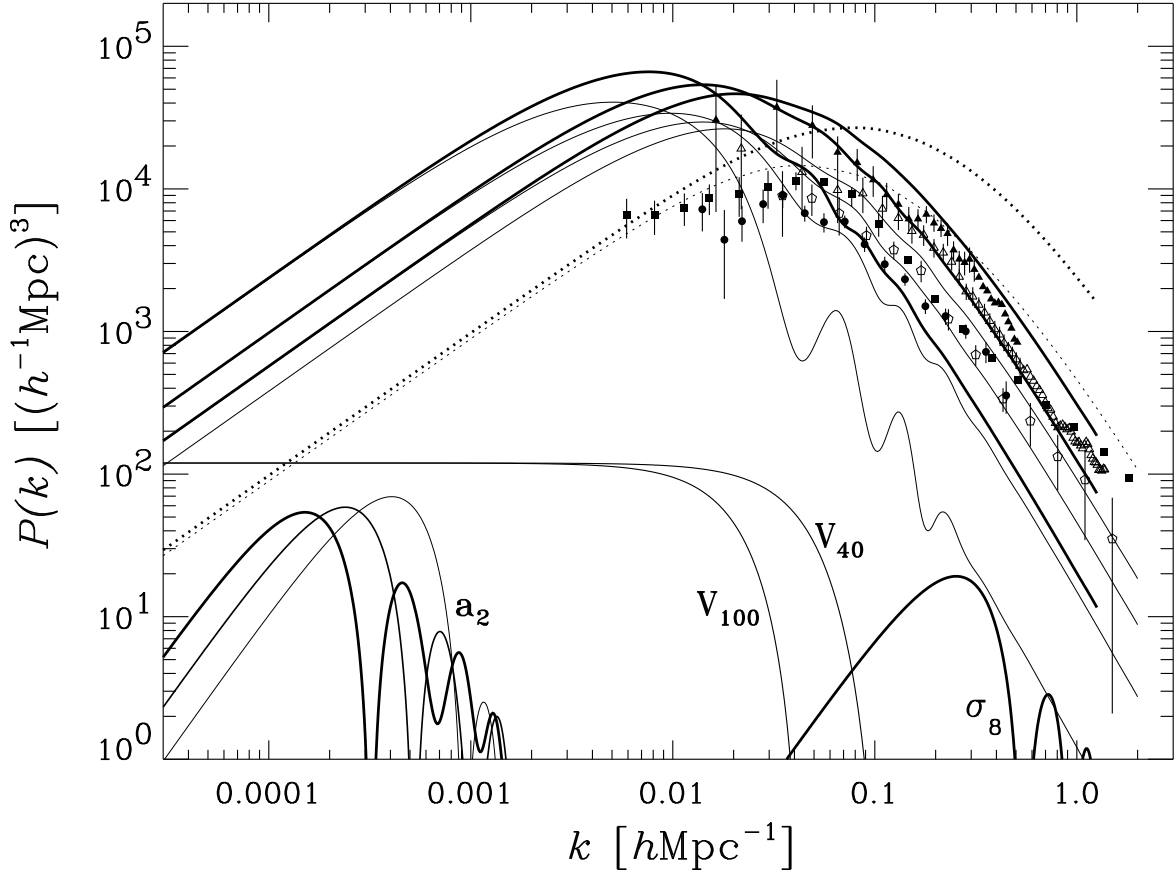

**Figure 6b** Same as figure 6a, but for non-zero cosmological constant models. Solid lines, in descending order, show the CDM-$\Lambda$ power spectra for $\Omega_0 = 0.1,\ 0.2,\ 0.3$ — heavy type, and $\Omega_0 = 0.1,\ 0.2,\ 0.3$, and $0.4$ — light type. The $h = 0.5$, $\Omega_b = 0.05$, and $h = 0.8$, $\Omega_b = 0.02$ models are shown in thin and heavy lines, respectively. Dotted lines correspond to $\Omega_0 = 1$, CDM models. Since the dependence of the fitted $Q_{rms-PS}$ amplitude on model parameters is rather weak, and because the SW and ISW effects scale with the total density, there is a strong resulting variation with $\Omega_0$ of the amplitudes of the matter power spectra, as shown in Fig. 4. Note that the quadrupole window function depends (due to the SW and ISW effects) on model parameters, $Q^2_{rms-PS} = (2\pi^2)^{-1} \int dk\, P(k) W_Q(k)$. For the pure Sachs-Wolfe effect, $W_Q(k) = \Omega_0^2 H_0^4 j_2^2(k R_{LSS})/4k^2$, where $R_{LSS}$ is the radius of the last scattering surface (at a redshift $z_{LSS} \simeq 1100$). The windows for $\Omega_0 = 0.1,\ 0.3$, and $1$ are shown from left to right (note the trend in the window maxima and changes in the shape of the functions).

The low-$\Omega_0$ CDM-$\Lambda$ (with $\Omega_0 \geq 0.15$) models can not be challenged solely on the basis of the $COBE$-DMR data, although the most likely $\Omega_0$ value is strikingly close to unity (see also Bunn & Sugiyama 1994). Conversely, observations of the matter distribution do require low values of the total density resulting in the constraint $\Omega_0 h^2 \lesssim 0.13$. On larger scales, these models predict bulk flows in reasonable agreement with POTENT for $\Omega_0 h \gtrsim 0.06$, but dramatically smaller than determined by the Lauer & Postman analysis for any ($\leq 1$) value of $\Omega_0$. Therefore, if the latter observation is confirmed all CDM-$\Lambda$ models (together with critical density CDM or MDM models) will be found wanting. The predicted mass distribution on scales of a few hundred Mpc overshoots most of the experimental estimates (including those derived from the IRAS 1.2Jy & APM surveys), but is in rough agreement with that recently obtained by da Costa et al. (1994).

Whilst it would be safer to await the final 4-year COBE results before offering definitive statements as to the viability of theoretical models, one should note that the CDM normalisation derived from the two year $COBE$-DMR data does appear to be irreconcilably high, while the MDM model has little room left for adjustment. Although the position of the CDM-$\Lambda$ models is more comfortable than that of critical density CDM models, it does not remain free from potentially fatal flaws. Unfortunately, CMB anisotropies on a one degree scale do not offer any serious prospects for distinguishing between cosmological constant models and other viable scenarios (Bond et al. 1994).

Krauss & Turner (1995) have recently proposed that cosmological constant CDM models with $\Omega_\Lambda$ in the range 0.6 – 0.7 can account for both large-scale structure and CMB observations. Similarly, Ostriker & Steinhardt (1995) have suggested that a spatially flat model with $\Omega_\Lambda = 0.65 \pm 0.1$ and a small tilt, $0.8 < n < 1.2$, is consistent with a number of cosmological constraints (including direct measures of $H_0$, stellar ages, large-scale structure statistics and CMB anisotropies). Accepting their challenge now to identify a problem for such models, we refer the reader again to fig. 6b. It appears that agreement may be achieved on both COBE and $\sigma_8$ scales. However, because the $b \geq 1$ condition requires a relatively low value of the Hubble constant, $h \lesssim (0.13/\Omega_0)^{1/2} \sim 0.65$ for $\Omega_0 \geq 0.3$ in the Harrison-Zel'dovich case, a small negative tilt ($n < 1$) may be necessary if $h$ is to be as high as $\sim 0.7 - 0.8$. Such a tilt would enhance the peak in the matter power spectrum at $k \sim 0.01\, h\mathrm{Mpc}^{-1}$ – where the current observational status remains uncertain – relative to the power on $\sigma_8$ scales near $k \sim 0.1\, h\mathrm{Mpc}^{-1}$. Therefore, definite conclusions will have to await more reliable large-scale, deep galaxy surveys capable of unraveling the shape of the galaxy power spectrum on scales up to comoving length-scales $\sim 600\, h^{-1}$Mpc.

## ACKNOWLEDGMENTS

We acknowledge the efforts of those contributing to the *COBE*-DMR. This work was supported in part by the Office of Space Sciences of NASA Headquarters. RS acknowledges the partial support of the Polish Scientific Committee (KBN) under grant no. 2P30401607 and of the Laboratory for Astronomy and Solar Physics and *COBE* project during his stay at GSFC, where part of this work was completed. We are grateful to L. da Costa, K. Fisher, M. Vogeley for providing their power spectra, and A. Dekel for providing the bulk flow points.